\documentclass[preprint]{raa}            % referee version: for submission

\usepackage{graphicx,times}             %for PS/EPS graphics inclusion, new
\usepackage{natbib}
\usepackage{amssymb,amsmath}
\bibpunct{(}{)}{;}{a}{}{,}
\footnotesize
\newdimen\digitwidth
\setbox0=\hbox{\rm0}
\digitwidth=\wd0
\catcode`!=\active
\def!{\kern\digitwidth}
\normalsize

\usepackage[pagebackref=true]{hyperref}

\begin{document}

\title{FAST Observations of Four Comets to Search for the Molecular Line Emissions between 1.0 and 1.5 GHz Frequencies}
%   \subtitle{I. Place Your Subtitle Here}

\volnopage{Vol.0 (20xx) No.0, 000--000}      %%preserved for Editor. DOn't remove!
\setcounter{page}{1}          %%starting page, preserved for Editor. DOn't remove!

\author{
Long-Fei Chen\inst{1,2,3}
\and Chao-Wei Tsai\inst{4,5,6}
\and Jian-Yang Li\inst{7}
\and Bin Yang\inst{8}
\and Di Li\inst{4,3,9}
\and Yan Duan\inst{10}
\and Chih-Hao Hsia\inst{11}
\and Zhichen Pan\inst{4}
\and Lei Qian\inst{4}
\and Donghui Quan\inst{3}
\and Xue-Jian Jiang\inst{3}
\and Xiaohu Li\inst{12}
\and Ruining Zhao\inst{13,14}
\and Pei Zuo\inst{4}
}

\institute{
School of Physics and Electronic Science, Guizhou Normal University, Guiyang 550025, China\\
\and Guizhou Provincial Key Laboratory of Radio Astronomy and Data Processing, Guiyang 550025, China\\
\and Research Center for Astronomical Computing, Zhejiang Laboratory, Hangzhou 311100, China\\
\and National Astronomical Observatories, Chinese Academy of Sciences, Beijing 100101, China; {\it cwtsai@nao.cas.cn}\\
\and Institute for Frontiers in Astronomy and Astrophysics, Beijing Normal University,  Beijing 102206, China\\
\and Key Laboratory of Radio Astronomy and Technology, Chinese Academy of Sciences, A20 Datun Road, Chaoyang District, Beijing, 100101, China\\
\and School of Atmospheric Sciences, Sun Yat-sen University, Zhuhai, Guangdong, China\\
\and N\'{u}cleo de Astronom\'{i}a, Facultad de Ingenier\'{i}ay Ciencias, Universidad Diego Portales, Chile\\
\and Department of Astronomy, College of Physics and Electronic Engineering, Qilu Normal University, 2 Wenbo Road, Zhangqiu District, Jinan 250200, China\\
\and Department of Electronic and Optical Engineering, Space Engineering University, Beijing, China\\
\and Laboratory for Space Research, Faculty of Science, The University of Hong Kong, Hong Kong (SAR), China\\
\and Xinjiang Astronomical Observatory, Chinese Academy of Sciences, No. 150 Science 1-Street, Urumqi 830011, China\\
\and CAS Key Laboratory of Optical Astronomy, National Astronomical Observatories, Chinese Academy of Sciences, Beijing 100101, China\\
\and School of Astronomy and Space Sciences, University of Chinese Academy of Sciences, Beijing 100049, China\\
\vs\no
{\small Received 20xx month day; accepted 20xx month day}}

\abstract{
We used the Five-hundred-meter Aperture Spherical radio Telescope (FAST) to search for the molecular emissions in the L-band between 1.0 and 1.5 GHz toward four comets, C/2020 F3 (NEOWISE), C/2020 R4 (ATLAS), C/2021 A1 (Leonard), and 67P/Churyumov-Gerasimenko during or after their perihelion passages. Thousands of molecular transition lines fall in this low-frequency range, many attributed to complex organic or prebiotic molecules. We conducted a blind search for the possible molecular lines in this frequency range in those comets and could not identify clear signals of molecular emissions in the data. Although several molecules have been detected at high frequencies of great than 100 GHz in comets, our results confirm that it is challenging to detect molecular transitions in the L-band frequency ranges. The non-detection of L-band molecular lines in the cometary environment could rule out the possibility of unusually strong lines, which could be caused by the masers or non-LTE effects. Although the line strengths are predicted to be weak, for FAST, using the ultra-wide bandwidth receiver and improving the radio frequency interference environments would enhance the detectability of those molecular transitions at low frequencies in the future.
}

\authorrunning{Chen et al.}            %author_head in even pages
\titlerunning{FAST Observations of Four Comets}  % title_head in odd pages

\maketitle
%% The author head (on even pages) and the title head (on odd pages) will be
%% automatically extracted from \author{} and \title{}. Whenever the title is too long,
%% you will be asked to supply a shorter one by inserting either \authorrunning{} or
%% \titlerunning{} before \maketitle. Anyway, you can specify your own heads.
%%
%%
%% Note: In the following text body of your manuscript, please note several differences from
%%       other major journals:
%% (1) \subsection{Please Capitalize the First Letter of Each Notional Word in Subsection Title}
%% (2) Please Capitalize the First Letter of Each Notional Word in all tables' captions

%
%________________________________________________ sections below
%
\section{Introduction}
Comets are considered ``fossils'' of our Solar System according to the planet formation theories \citep{Shu1977,Brasser2013}. Following the formation of the Solar Nebula from molecular cloud to protostar and protoplanetary disk, planetesimals formed and finally evolved to the present day Solar System \citep{Ehrenfreund2000,Chambers2023}. Thus, the composition of comets may reflect the initial environments of our Solar System \citep{vanDishoeck2013,Willacy2022}. For example, the various molecular species detected in the cometary coma may have inherited from the protosolar nebula \citep{Bergner2021}, where they are frozen onto and preserved in the ice mantles of the dust grains until the comets travel to inner Solar System to the Sun to release the species into the gas phase under solar heating.

The composition of comets offers a useful chemical way to study the molecular inheritance from the Solar Nebula's evolutionary history. More than 70 molecules have been identified from various comets by remote or in-situ measurements \citep{Rubin2019}, with 67P/Churyumov-Gerasimenko having the highest number of discovered molecules \citep{Biver2019} from measurements by the Rosetta mission, including both simple species such as hydroxyl \citep{Smith2021}, and complex species such as glycine \citep{Altwegg2016}.

Besides the identification of molecules by in situ mass spectrometer measurements and by electron excitation in ultraviolet and visible wave ranges, the rotationally excited transitions of the gas-phase molecules located in the (sub-)millimeter frequency range provide the best way to detect them remotely by radio observations to reveal the kinetics and excitation conditions of the environment where these molecules are located. For example, using Atacama Large Millimeter/Submillimeter Array (ALMA), \citet{Roth2021} detected the emissions from methanol, formaldehyde, and other species from comet C/2015 ER61 (PanSTARRS) in the radio frequency around $\sim$350 GHz, revealing the low coma kinetic temperature and its asymmetric expansion velocity. \citet{Bergman2022} detected the methanol and hydrogen cyanide in several comets using the Onsala 20-m telescope at a frequency near 90 GHz, revealing short-time activity variations of those comets.

Although many observations of molecular transitions have been reported in the 10 GHz and higher frequency range, the lower frequency range is under-explored, and the molecules detected in these bands only represent a small fraction of the molecules detected in comets. Detection of molecular transitions in the low-frequency ranges is challenging. \citet{Salter2008} conducted a line search within a bandwidth between 1.1 and 10 GHz using the Arecibo Telescope toward the starburst galaxy, Arp 220. Several absorption lines were reported, with their rest frequencies all above 4 GHz. Only one absorption feature around 1.638 GHz was found, which was attributed to $^{18}$OH or formic acid (HCOOH) due to line confusion. \citet{Tan2020} performed surveys toward star-forming regions to search for molecular lines in the frequency range 6.0--7.4 GHz using the same telescope, resulting in the detection of only three molecules, CH$_3$OH, H$_2$CS, and OH. In the low frequencies down to 1 GHz of the radio L-band, only the OH 18 cm $\Lambda$-doublet transitions have been detected for many comets using the Arecibo Telescope, Nan\c{c}ay Radio Telescope, and Green Bank Telescope \citep{Lovell2002, Tseng2007, Smith2021, Drozdovskaya2023}. At even lower frequencies around 0.1 GHz, tentative detection of NO and its isotopologues, t-DCOOH, $^{17}$OO, as well as SH were reported toward Galactic Centre and Orion using the Murchison Widefield Array \citep{Tremblay2017, Tremblay2018, Tremblay2020}.

The Five-hundred-meter Aperture Spherical radio Telescope (FAST) is the world's largest single-dish radio telescope with three times the sensitivity of the Arecibo Telescope \citep{Nan2011, Li2018, Qian2020}. Thus, it provides a great opportunity to search for molecules in the frequency range from 1.0 to 1.5 GHz with its 19-beam receiver. In this paper, we report the first attempt to search for molecular lines in four comets observed by FAST in 2020 and 2021. The observational setups are presented in Section \ref{sec.obs}. Section \ref{sec.res} and \ref{sec.dis} discuss the results and the interpretations, respectively. We reach our conclusions in Section \ref{sec.sum}.

\section{Observations}\label{sec.obs}
Four comets were observed from August 2020 to December 2021, including three long-period comets, C/2020 F3 (NEOWISE) (hereafter NEOWISE, \citep{Bauer2020, Biver2022}), C/2020 R4 (ATLAS) \citep{Manzini2021}, and C/2021 A1 (Leonard) \citep{Leonard2021, Zhang2021}, and one Jupiter-family comet, 67P/Churyumov-Gerasimenko (hereafter 67P/C-G) \citep{Biver2023}. Table \ref{tab.comets} summarizes the observation log. The Ephemeris of each comet was obtained from the HORIZONS system of Jet Propulsion Laboratory \citep{Giorgini1996}. All comets were moving at less than 15\arcmin~per minute during the tracking, well within the 30\arcmin~per minute tracking limit of FAST \citep{Jiang2020}. The observations of comet NEOWISE were conducted during the gap between two maintenance periods of FAST when the comet was near its closest approach to Earth, and the observations of other comets were Target of Opportunity (ToO) observations. Table \ref{tab.setups} lists the FAST programs to observe these comets.

For comet NEOWISE, the target was observed with 12 observation blocks, each with an on-source integration time of 10 minutes in a tracking observation mode centered in the central beam (M01) of the FAST 19-beam receiver. To obtain off-source measurement for baseline subtraction, every 4 on-source observation schedules were accompanied by an observation schedule at $\sim$3 degrees from the target. Pulsar and spectral data were simultaneously recorded in the FAST backend during the observations. The pulsar data were used to search for pulsar candidates with no positive detections \citep{Pan2021, Qian2021}. The spectral data were used in this work to search for molecular lines from the comet. It should be noted that for the tracking mode used for comet NEOWISE, the comet was actually not tracked. The telescope pointing was targeted at a fixed RA and Dec coordinate that was consistent with the position of the comet NEOWISE during each observation block. However, due to the moving speed is faster for comet than for background stars, the comet would move out of the beam after a period of time. Thus, the tracking time for comet NEOWISE was limited to 10 minutes at each observation block to ensure the comet remained within the 2\farcm9 (at 1.4 GHz) beam size of M01. The left panel of Figure \ref{fig.neowise_C2020R4} shows the optical image of comet NEOWISE taken at the site of FAST during the observations.

\begin{figure}
\centering
\includegraphics[scale=0.5]{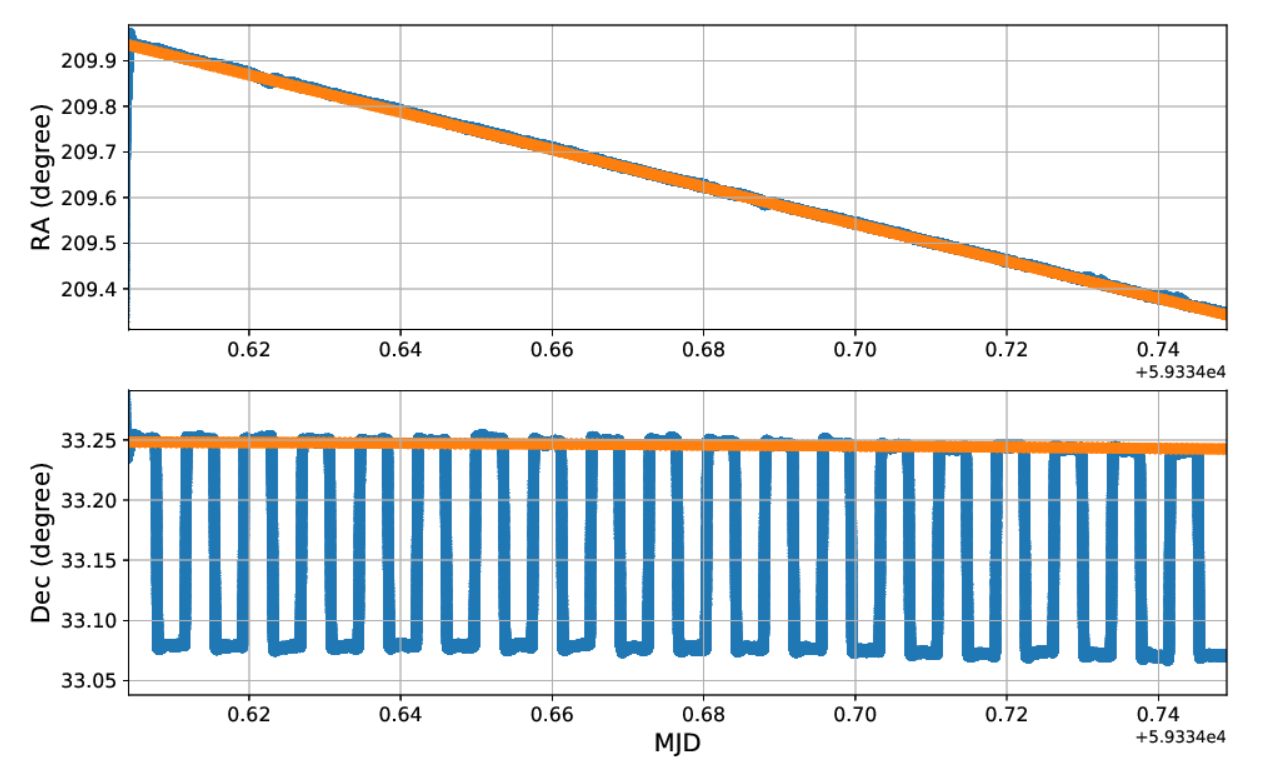}
\caption{The telescope tracking (blue) and the actual RA and Dec coordinates (orange) for comet C/2020 R4 (ATLAS) on 30 April 2021.
}
\label{fig.C2020R4_tracking}
\end{figure}

\begin{figure}
\centering
\includegraphics[scale=0.3]{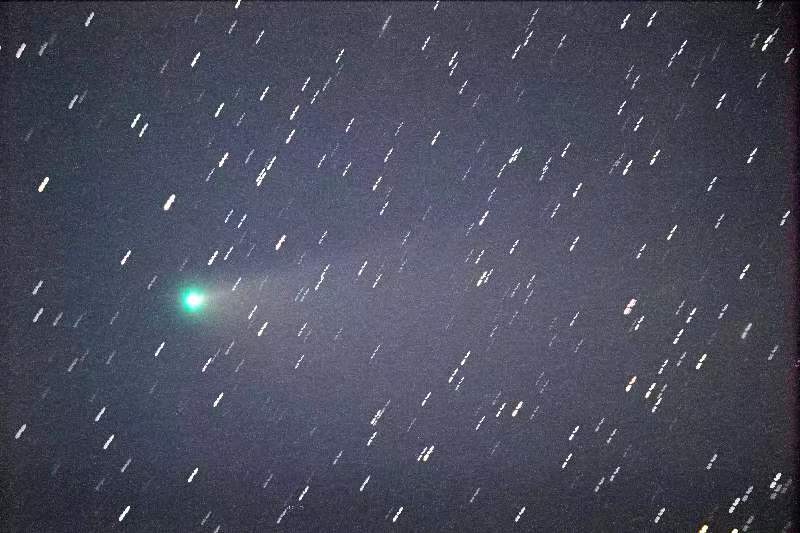}
\includegraphics[scale=0.21]{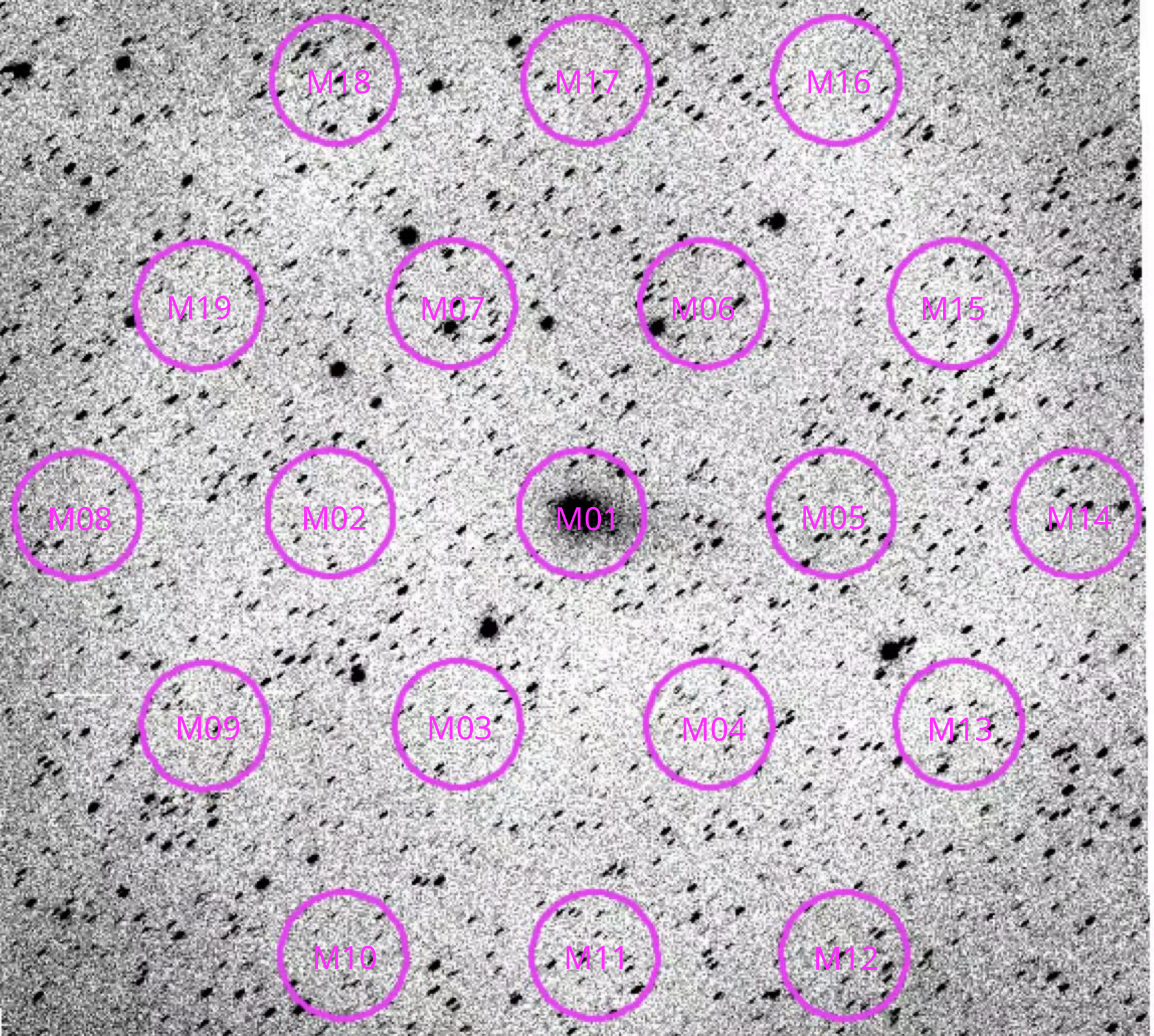}
\caption{(a) The image of comet NEOWISE stacked with 20 pictures, each with an exposure of 30 seconds. It was taken with a 127 mm refraction telescope at the FAST site on 30 July 2020. (b) The $25\arcmin \times 23\arcmin$ optical image of comet C/2020 R4 (ATLAS) from Lulin observatory taken on 30 April 2021 overlaid with the beam positions of the FAST 19-beam receiver.
}
\label{fig.neowise_C2020R4}
\end{figure}

For the other three comets that we observed, a custom observation mode was developed in order to improve the observational efficiency for moving targets. Compared with the tracking mode used for comet NEOWISE, these three comets were tracked at their non-sidereal rates in this custom mode, i.e.~their RA and Dec coordinates were update in real time. In this mode, the target comet was placed at the central beam M01 and the side beam M11 of the 19-beam receiver alternatively with a 5-minute integration in each beam. This new mode would allow one beam to acquire the source signal while the other simultaneously acquires the off-source sky background. When alternating between two beams, the source can be continuously observed, although the sky subtraction has to be performed for each beam individually because different beams have different gains. For our observations, the separation of 10\farcm22 between M01 and M11 required 30 seconds to switch between them to avoid exceeding the acceleration limit of 90\arcsec ~per second for the FAST receiver maneuver. Figure \ref{fig.C2020R4_tracking} shows the tracking positions and the orbital coordinates of comet C/2020 R4 (ATLAS) on 30th April 2021. The right panel of Figure \ref{fig.neowise_C2020R4} shows its optical image centered at the M01 beam and overlaid by the FAST 19-beam receiver.

Table \ref{tab.setups} summarizes the observation setups. The spectral backend with 1,048,576 channels was configured to cover 1.0 to 1.5 GHz at a frequency resolution of 476.8 Hz, corresponding to a velocity resolution of 0.1 km/s at 1.420 GHz. The sampling time was 1 second for the observations on comet NEOWISE and 0.1 second for other comets. The flux calibration utilized the noise diode with an injection to the optical path in the low-intensity mode (1 K) for NEOWISE and the high-intensity mode (10 K) for other comets. The noise injection period varied for different comets, as listed in Table \ref{tab.setups}. Assuming a system temperature of 24 K \citep{Jiang2020} and an hour of integration, the rms noise level could reach 13 mK.

\begin{table}
\begin{center}
\caption[]{Observing conditions of the four comets observed by FAST.}
\label{tab.comets}
\begin{tabular}{llllll}
\hline\noalign{\smallskip}
Comet name           & Start observing  & Magnitude &Heliocentric   &Geocentric    & $\dot{\Delta}$$^a$\\
                     & UT time          & (mag)     & distance (AU) &distance (AU) & (km/s)\\
\hline\noalign{\smallskip}
C/2020 F3 (NEOWISE)  & 2020-08-01 06:15 & 5.9       &0.827          &0.793         & 35.84\\
C/2020 R4 (ATLAS)    & 2021-03-31 23:30 & 13.9      &1.138          &0.947         & -49.86\\
C/2020 R4 (ATLAS)    & 2021-04-02 23:30 & 13.9      &1.152          &0.890         & -49.31\\
C/2020 R4 (ATLAS)    & 2021-04-29 14:30 & 13.5      &1.400          &0.506         & 27.80 \\
C/2020 R4 (ATLAS)    & 2021-04-30 14:30 & 13.6      &1.411          &0.522         & 31.37 \\
67P/C-G              & 2021-10-18 20:30 & 10.1      &1.225          &0.442         & -3.74 \\
67P/C-G              & 2021-11-02 19:30 & 10.0      &1.211          &0.421         & -1.26  \\
C/2021 A1 (Leonard)  & 2021-12-07 00:40 & 5.3       &0.846          &0.326         & -49.00\\
% 29P                & 2021-10-05       & & & & 120  & no noise injection
\hline\noalign{\smallskip}
\end{tabular}
\end{center}
\tablecomments{0.86\textwidth}{$^a$ Line-of-sight velocity of the target with respect to the observer. A positive value means the target center is moving away from the observer, negative indicates movement toward the observer. Data were taken from JPL/Horizons ephemeris.}
\end{table}

\begin{table}
\begin{center}
\caption[]{Receiver configurations for our observations of the four comets.}
\label{tab.setups}
\begin{tabular}{llll}
\hline\noalign{\smallskip}
Comet name           & Duration   & Noise injection period      & Program ID \\
                     & (minutes)  & (seconds, cal-on + cal-off) &  \\
\hline\noalign{\smallskip}
C/2020 F3 (NEOWISE)  & 120        & 2 (1 + 1)                   & - \\
C/2020 R4 (ATLAS)    & 65         & 10 (0.1 + 9.9)              & PT2020\_0166 \\
C/2020 R4 (ATLAS)    & 65         & 10 (0.1 + 9.9)              & PT2020\_0166 \\
C/2020 R4 (ATLAS)    & 220        & 10 (0.1 + 9.9)              & PT2020\_0166 \\
C/2020 R4 (ATLAS)    & 210        & 10 (0.1 + 9.9)              & PT2020\_0166 \\
67P/C-G              & 120        & 8 (1 + 7)                   & PT2021\_0045 \\
67P/C-G              & 90         & 8 (1 + 7)                   & PT2021\_0045 \\
C/2021 A1 (Leonard)  & 120        & 8 (1 + 7)                   & PT2021\_0045 \\
\hline\noalign{\smallskip}
\end{tabular}
\end{center}
\tablecomments{0.86\textwidth}{The dates of observations for these comets are in Table \ref{tab.comets}. The PIs of these observations are Zhichen Pan for comet NEOWISE, Chao-Wei Tsai for program ID PT2020\_0166, and Zhong-Yi Lin for program ID PT2021\_0045, respectively.}
\end{table}

\section{Data processing and analysis}\label{sec.res}
The receiver recorded the full polarization of the signal in the backend during the observations. We first averaged the total power in the two perpendicular polarization directions, and converted the total power to antenna temperature using Equation \ref{eq.ta},

\begin{equation}
T_{\rm a} = T_{\rm noise} \times \frac{\mathrm{Power_{cal-off}}}{\mathrm{Power_{cal-on} - Power_{cal-off}}},
\label{eq.ta}
\end{equation}

where $T_{\rm noise}$ is the temperature of noise injection, and $\mathrm{Power_{\rm cal-on}}$ and $\mathrm{Power_{\rm cal-off}}$ are the total power when the noise injection was switched on and off, respectively. From the spectra for CH$_3$OH (Figure \ref{fig.CH3OH_67P}), we can see that the rms of the final spectrum is consistent with the estimated noise level.

\subsection{Molecular line searching}
To search for the molecular transition lines in the L-band, we collected a list of lines from the Splatalogue\footnote{https://splatalogue.online/} database in the frequency range from 1.0 to 1.5 GHz. We carried out a systematic blind search based on the prospective molecular lines in selected frequency ranges without strong radio frequency interference (RFI). Table \ref{tab.lines} shows examples of the prospective molecular transition lines for CH$_3$OH and $^{17}$OH in the L-band.

\begin{figure}
\centering
\includegraphics[width=\textwidth]{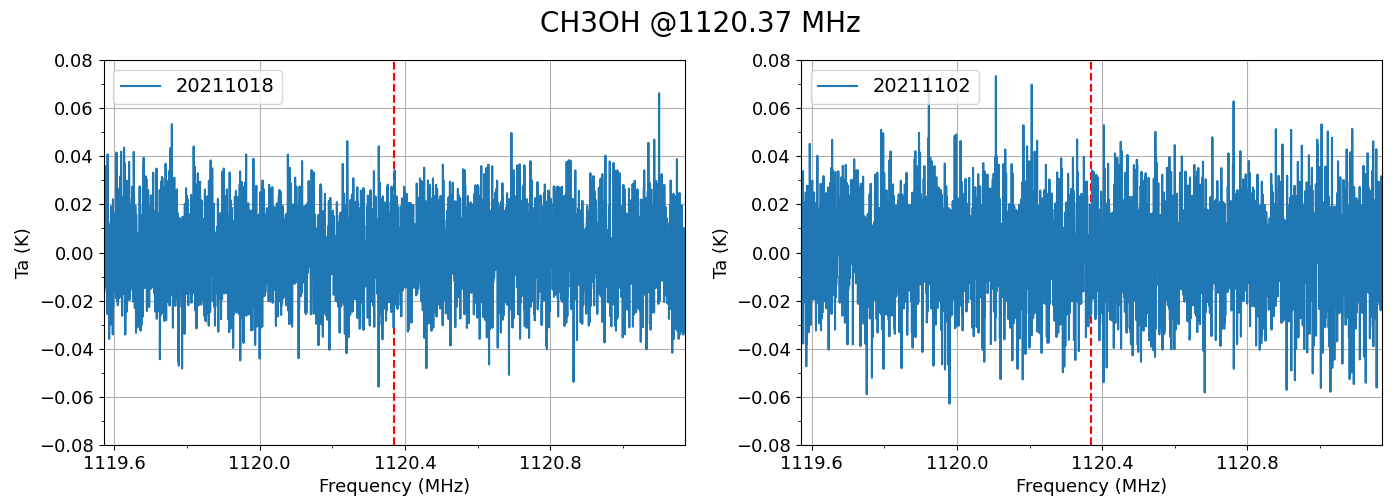}
\includegraphics[width=\textwidth]{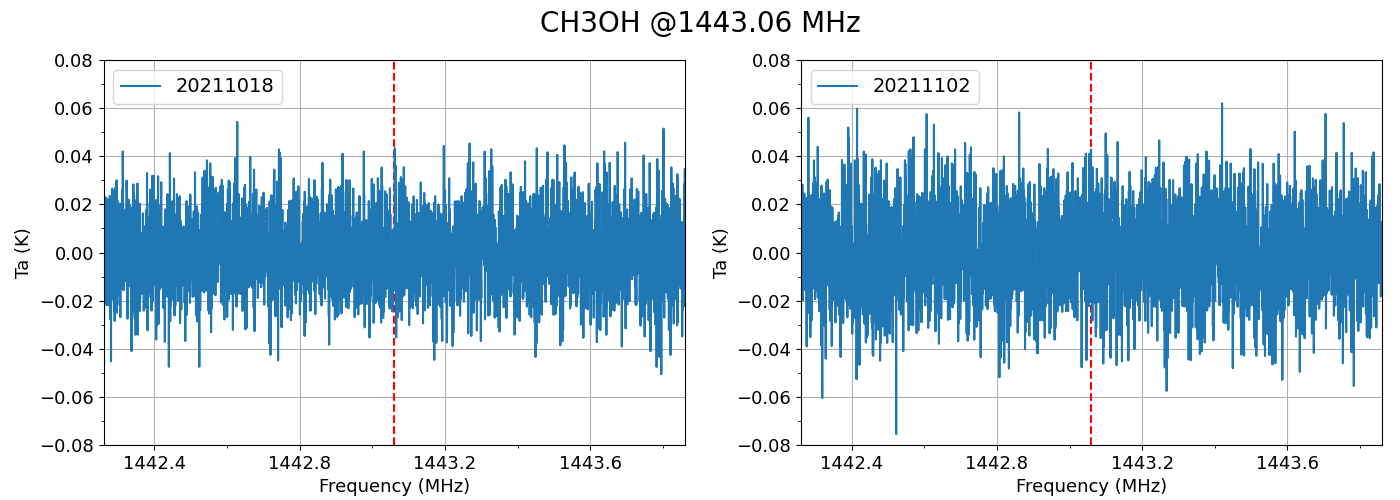}
\caption{The spectra of comet 67P/C-G obtained on 2021-10-18 and 2021-11-02. The red vertical dashed lines mark the rest frequencies of CH$_3$OH. The upper and lower panels are for CH$_3$OH at 1.120 and 1.443 GHz, respectively.
}
\label{fig.CH3OH_67P}
\end{figure}

\begin{figure}
\centering
\includegraphics[scale=0.4]{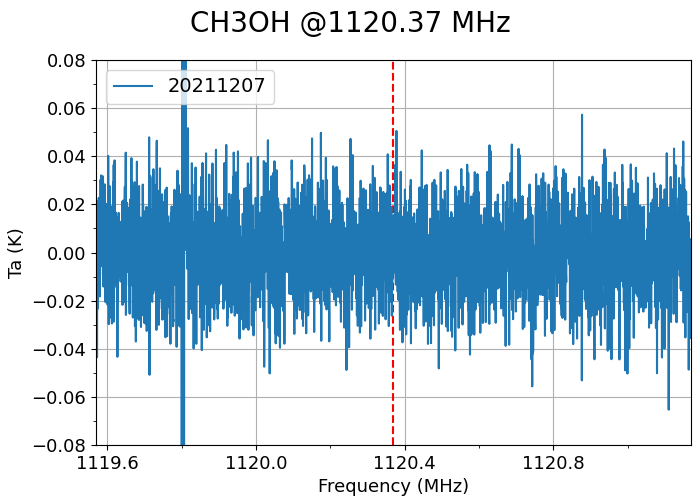}
\includegraphics[scale=0.4]{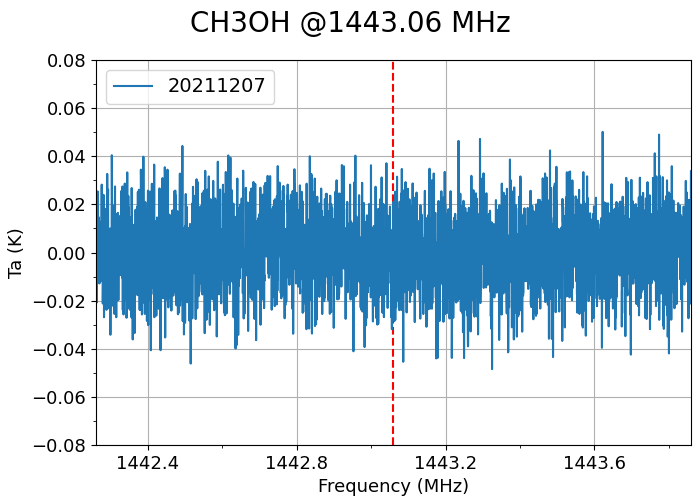}
\caption{The spectra of comet C/2021 A1 (Leonard) obtained on 2021-12-07. The red vertical dashed lines mark the rest frequencies of CH$_3$OH. The upper and lower panels for CH$_3$OH at 1.120 and 1.443 GHz, respectively.
}
\label{fig.CH3OH_C2021A1}
\end{figure}

Methanol is one of the common ice components and has been detected in a number of comets \citep{Rubin2019}. Figures \ref{fig.CH3OH_67P} and \ref{fig.CH3OH_C2021A1} show our spectra that cover the expected CH$_3$OH lines for comets 67P/C-G and C/2021 A1 (Leonard), respectively. However, no clear methanol emissions are discernible. A suspicious signal of CH$_3$OH at 1.120 GHz for comet C/2021 A1 (Leonard) is visible in Figure \ref{fig.CH3OH_C2021A1}. However, it is present in both the on-source and off-source observations for both beam M01 and beam M11 and therefore is most likely RFI. Assuming a rotation temperature of 36 K and a line width of 0.7 km/s for the methanol emission as indicated by the IRAM 30-m observations \citep{Biver2023}, the estimated 3$\sigma$ upper limit production rate of CH$_3$OH at 1.443 GHz for 67P/C-G is 2 $\times$ 10$^{32}$ molecules s$^{-1}$ for two hours of integration time using the equation from \citet{Drahus2010}. Compared with the derived production rate of CH$_3$OH (around 10$^{26}$ molecules s$^{-1}$) from \citet{Biver2023}, this estimated upper limit is too high to give a reasonable constrains. Due to the non-detection of molecular lines, we therefore constrained the integrated intensity of CH$_3$OH at 1.443 GHz based on its reported production rate and described the results in the later section.

\begin{figure}
\centering
\includegraphics[width=\textwidth]{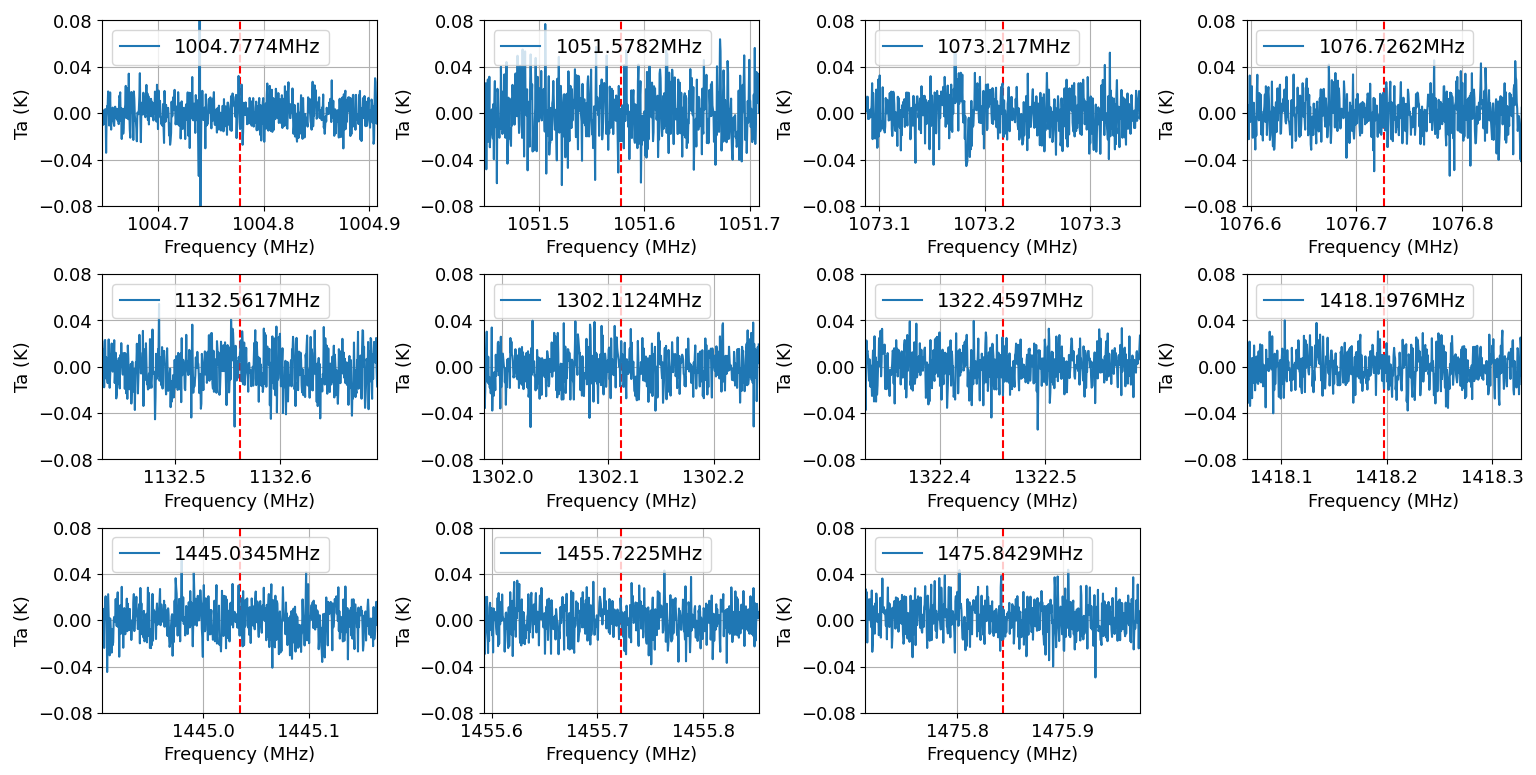}
\caption{The $^{17}$OH spectra of comet C/2021 A1 (Leonard) obtained on 2021-12-07. The red vertical dashed
lines mark the rest frequencies of $^{17}$OH.
}
\label{fig.17OH_C2021A1}
\end{figure}

The 18 cm OH line has been detected in many comets in the L-band by the Arecibo Telescope, Nan\c{c}ay Radio Telescope, and Green Bank Telescope \citep{Lovell2002, Tseng2007, Smith2021, Drozdovskaya2023}. Unfortunately, the OH transitions at 1.665 and 1.667 GHz are out of the frequency range of the FAST L-band receiver. However, several tens of $^{17}$OH transitions and six $^{18}$OH transitions are located within the FAST L-band receiver. Figure \ref{fig.17OH_C2021A1} shows those spectra for comet C/2021 A1 (Leonard). Again, no $^{17}$OH and $^{18}$OH transitions are visible in our data. However, there is a suspicious absorption feature near the expected transitions of 1073.214 MHz. Further inspection of the data suggests that this ``absorption'' resulted from the subtraction of the off-source spectrum from the on-source spectrum, which means that it is an emission both in the on-source and off-source spectrum. Therefore, we conclude that this is an artifact due to RFI.

No observations were reported for the detection of $^{17}$OH in comets in the literature. To estimate the upper limit column density of $^{17}$OH, we assumed a line width of 2.5 km/s based on the OH line profile for comet NEOWISE observed by Arecibo \citep{Smith2021}. Since the excitation temperature of $^{17}$OH in comet comae is unknown, we simply adopted a range of values from 10--100 K and calculated the 3$\sigma$ upper limit column density of $^{17}$OH at 1.302 GHz. It should be noted that the excitation temperature used here is not physically appropriate since the maser effects due to the absorption of solar radiation are very important for the OH excitation \citep{Schleicher1988}. Therefore, the above constrains on the $^{17}$OH column density are very crude. We found that the upper limit column density for $^{17}$OH is in the range of 7--10 $\times$ 10$^{12}$ cm$^{-2}$. The estimated value is about 2 times lower than the $^{16}$OH column density previously reported \citep{Smith2021}.

In the cometary environments, the production of OH is dominated by the photodissociation of H$_2$O. The molecule $^{17}$OH would be produced by the photodissociation of the isotopologue of water, H$_2^{17}$O. The oxygen isotopic abundance ratio of $^{17}$O/$^{16}$O as derived from H$_2^{17}$O/H$_2^{16}$O in comet 67P/C-G is $\sim$4$\times$10$^{-4}$ \citep{Altwegg2015, Muller2022}. If we assume the same abundance ratio of $^{17}$OH/$^{16}$OH for comet NEOWISE and 67P/C-G, then that would indicate that our upper limit column density of $^{17}$OH is highly overestimated. This suggests that the integrated intensity of $^{17}$OH used in the calculations, i.e., ~the rms noise level with the current integration time is too high to constrain the signal strength of $^{17}$OH. This also indicated that the non-detection of $^{17}$OH is consistent with the expected $^{17}$O/$^{16}$O elemental ratio.

Instead of constraining the molecular column density or production rate, we can use the molecular production rate reported in the literature to estimate the expected integrated intensity of L-band molecular emissions. The molecular production rate can be expressed by the following Equation \ref{eq.q}, assuming optically thin and LTE \citep{Drahus2010},

\begin{equation}
Q = \frac{2}{\sqrt{\pi ln2}} \frac{k_B}{h} \frac{b\Delta v_{exp}}{D I(T) \nu} (e^{h\nu/k_B T}-1) \int T_{mb}dv,
\label{eq.q}
\end{equation}

where $Q$ is the production rate in molecules per second, $k_B$ and $h$ are Boltzmann constant and Planck constant, respectively, $b$ = 1.2 is the dimensionless factor of the full width at half maximum (FWHM) of the telescope’s beam, $D$ is the antenna's aperture, $\Delta$ is the geocentric distance of the comet, $v_{exp}$ is the expansion velocity of gas, $\nu$ and $I(T)$ are the rest frequency of the molecule and line intensity at temperature $T$, respectively, and $\int T_{mb}dv$ is the integrated intensity. Using the production rates of molecules reported for comets in the literature, we can derive the intensity of molecular emissions. Table \ref{tab.intensity} shows the estimated integrated intensity for CH$_3$OH at 1.443 GHz and $^{17}$OH at 1.302 GHz for comets NEOWISE, 67P/C-G, and C/2021 A1 (Leonard), which have the production rates for CH$_3$OH and OH previously reported. It should also be noted here that these estimations are based on the LTE assumption, so the results are rough estimate. The maser effects or non-LTE assumption should be considered if one wants to obtain a more accurate estimate. The integrated intensity for OH at 1.667 GHz is also estimated for comparison. The estimated integrated intensity for CH$_3$OH at 1.443 GHz and $^{17}$OH at 1.302 GHz are orders of magnitude lower than that for OH at 1.667 GHz. Therefore, we do not expect the detection of those molecular lines in the L-band in our data, at least for CH$_3$OH and $^{17}$OH. The possible reasons are discussed in the following section.

Except for CH$_3$OH and $^{17}$OH, there are various other molecules whose transition lines fall in the L-band.
Table \ref{tab.molecules} lists a subset of molecules of interest. None of these molecules were identified, and their detection thresholds are far below the limit of current constrains. Several reasons can lead to the non-detection. Apart from CH$_3$OH or H$_2$CO, whose abundances are a few to ten percent relative to water, for other molecules, such as C$_2$H$_5$OH shown in the list, their abundances relative to water could be lower than one percent. On the other hand, large molecules tend to have more transition lines. However, due to the energy distribution in these lines, the intensity for transition lines at L-band is weaker than the high frequency lines. And also the non-LTE effects for the molecule excitation in the cometary coma environment may also complicate the situation \citep{Bockelee-Morvan2004}.

Finally, we examined the potential line features of CH$_3$OH, $^{17}$OH, and other molecules for the comets C/2020 R4 (ATLAS) and C/2020 F3 (NEOWISE). However, the RFI was more severe in the data obtained before November 2021 because the electronic environment of the FAST feed cabin was not improved until after that time. Therefore, fewer RFI-free bands in the 1.0--1.5 GHz were available for these two comets. In summary, we did not identify convincing molecular emission signals in the observations for these comets, either.

\begin{table}
\begin{center}
\caption[]{Molecular transition lines and their spectral parameters for CH$_3$OH and $^{17}$OH in the L-band.}
\label{tab.lines}
\begin{tabular}{p{40pt}p{120pt}p{60pt}p{40pt}p{40pt}p{40pt}}
\hline\noalign{\smallskip}
Molecule   & Quantum Numbers           & Rest frequency & log$_{10}$A$_{ij}$ & E$_{up}$  & S$_{ij}\mu^2$  \\
           & (lower - upper)           & (MHz)          & (s$^{-1}$)         & (K)       & (Debye$^2$)      \\
\hline\noalign{\smallskip}
CH$_3$OH   & 10(2,8) - 10(2,9) A, vt=0 & 1120.3700      & -12.35144          & 165.40152 & 2.28465  \\
           & 5(1,4) - 5(1,5) A, vt=1   & 1443.0600      & -12.04136          & 360.02069 & 1.14368  \\
\hline
$^{17}$OH  & N = 5 - 5, J+1/2 = 9/2 - 9/2, p = 1 - -1, F$_1$ = 4 - 4, F+1/2 = 3/2 - 3/2  & 1004.7774  & -12.39136  & 872.65326  & 0.13758 \\
           & N = 5 - 5, J+1/2 = 9/2 - 9/2, p = 1 - -1, F$_1$ = 5 - 5, F+1/2 = 5/2 - 5/2  & 1051.5782  & -12.33068  & 872.65681  & 0.20702 \\
           & N = 1 - 1, J+1/2 = 3/2 - 3/2, p = -1 - 1, F$_1$ = 1 - 2, F+1/2 = 5/2 - 3/2  & 1073.217   & -13.76028  & 0.10158    & 0.00724 \\
           & N = 5 - 5, J+1/2 = 9/2 - 9/2, p = 1 - -1, F$_1$ = 5 - 4, F+1/2 = 5/2 - 3/2  & 1076.7262  & -13.60391  & 872.65672  & 0.01028 \\
           & N = 1 - 1, J+1/2 = 3/2 - 3/2, p = -1 - 1, F$_1$ = 1 - 2, F+1/2 = 7/2 - 7/2  & 1132.5617  & -12.50549  & 0.07795    & 0.14774 \\
           & N = 1 - 1, J+1/2 = 3/2 - 3/2, p = -1 - 1, F$_1$ = 1 - 1, F+1/2 = 5/2 - 3/2  & 1302.1124  & -10.81113  & 0.10163    & 3.60695 \\
           & N = 1 - 1, J+1/2 = 3/2 - 3/2, p = -1 - 1, F$_1$ = 2 - 2, F+1/2 = 7/2 - 5/2  & 1322.4597  & -10.77313  & 0.10318    & 5.01041 \\
           & N = 1 - 1, J+1/2 = 3/2 - 3/2, p = -1 - 1, F$_1$ = 1 - 2, F+1/2 = 3/2 - 1/2  & 1418.1976  & -10.78143  & 0.11943    & 1.99290 \\
           & N = 1 - 1, J+1/2 = 3/2 - 3/2, p = -1 - 1, F$_1$ = 1 - 2, F+1/2 = 3/2 - 3/2  & 1445.0345  & -11.22889  & 0.11942    & 0.67236 \\
           & N = 1 - 1, J+1/2 = 3/2 - 3/2, p = -1 - 1, F$_1$ = 2 - 2, F+1/2 = 5/2 - 3/2  & 1455.7225  & -10.62898  & 0.11993    & 3.92646 \\
           & N = 1 - 1, J+1/2 = 3/2 - 3/2, p = -1 - 1, F$_1$ = 1 - 2, F+1/2 = 3/2 - 1/2  & 1475.8429  & -11.52867  & 5125.78645 & 0.31647 \\
\hline\noalign{\smallskip}
\end{tabular}
\end{center}
\end{table}

\begin{table}
\begin{center}
\caption[]{The estimated integrated intensity for molecular lines of CH$_3$OH (at 1.443 GHz), $^{17}$OH (at 1.302 GHz), and OH (at 1.667 GHz) for comets C/2020 F3 (NEOWISE), 67P/C-G, and C/2021 A1 (Leonard).}
\label{tab.intensity}
\begin{tabular}{llll}
\hline\noalign{\smallskip}
Comet name           & Molecule  & Production rate                                & Estimated integrated intensity\\
                     &           & (molecules s$^{-1}$)                           & (mK km/s)\\
\hline\noalign{\smallskip}
C/2020 F3 (NEOWISE)  & CH$_3$OH  & 1.25$\times$10$^{28}$ \citep{Biver2022}      & 1.6$\times$10$^{-4}$ \\
                     & $^{17}$OH & 1.44$\times$10$^{25}$ $^a$                   & 1.7$\times$10$^{-4}$ \\
                     & OH        & 3.6$\times$10$^{28}$ \citep{Smith2021}       & 8.1 \\
\hline
67P/C-G              & CH$_3$OH  & 3.67$\times$10$^{26}$ \citep{Biver2023}      & 6.4$\times$10$^{-6}$ \\
                     & $^{17}$OH & 4.4$\times$10$^{24}$ $^a$                    & 4.9$\times$10$^{-5}$ \\
                     & OH        & 1.1$\times$10$^{28}$ \citep{Biver2023}       & 2.4 \\
\hline
C/2021 A1 (Leonard)  & $^{17}$OH & 3.7$\times$10$^{25}$ $^a$                    & 1.9$\times$10$^{-3}$ \\
                     & OH        & 9.31$\times$10$^{28}$ \citep{Skirmante2022}  & 94.6 \\
\hline\noalign{\smallskip}
\end{tabular}
\end{center}
\tablecomments{0.86\textwidth}{
$^a$ Assuming Q($^{17}$OH) = 4$\times$10$^{-4}$ Q(OH).}
\end{table}

\begin{table}
\begin{center}
\caption[]{A subset of molecules of interest used for searching in this study.}
\label{tab.molecules}
\begin{tabular}{p{120pt}p{250pt}}
\hline\noalign{\smallskip}
Groups                    & Molecules  \\
\hline
P-bearing molecules       & PO, PH$_3$ \\
\hline
S-bearing molecules       & SH, H$_2$CS, H$_2$SO$_4$ \\
\hline
Carbon-chains             & C$_6$H, C$_9$H, C$_6$O, C$_7$O, C$_9$O, HC$_7$N, HC$_9$N, HC$_{11}$N \\
\hline
Complex organic molecules & CH$_3$OH, CH$_3$CHO, CH$_3$COOH, CH$_3$CH$_2$CHO, C$_2$H$_5$OH, C$_2$H$_5$CN, NH$_2$CH$_2$CH$_2$OH, H$_2$NCH$_2$COOH-I, H$_2$NCH$_2$COOH-II, a'GG'g-CH$_2$OHCH$_2$CH$_2$OH, aG'g-CH$_3$CHOHCH$_2$OH \\
\hline
Others                    & $^{17}$OH, $^{18}$OH, OD, H$_2$CO, D$_2$CO, NO, NH$_3$, ND$_3$ \\
\hline\noalign{\smallskip}
\end{tabular}
\end{center}
\end{table}

\subsection{The neutral hydrogen line}
Since our primary scientific goal is to search the molecular lines, we did not specifically consider the background contamination of comets when scheduling our observation plans, which is important if we want to identify the HI emission from the comet itself. We have visually checked the background contamination by stacking the comets position and the HI4PI survey \citep{HI4PI}. We found that, for comet NEOWISE and the last two scheduled observations of comet C/2020 R4 (ATLAS), the background was clean. While for other comets during their observations, they were either coincide with a background cloud or on the edge of a background cloud, thus unsuitable for constraining the HI emission from those comets. As a result, the HI spectrum from comet 67P/C-G, C/2021 A1 (Leonard), and C/2020 R4 (ALTAS) have a complex profile with multiple components, likely due to the background contamination. Most importantly, the off-source positions were also characterised by similar HI spectrum, thus we were also unable to perform a good baseline subtraction.

\begin{figure}
\centering
\includegraphics[scale=0.5]{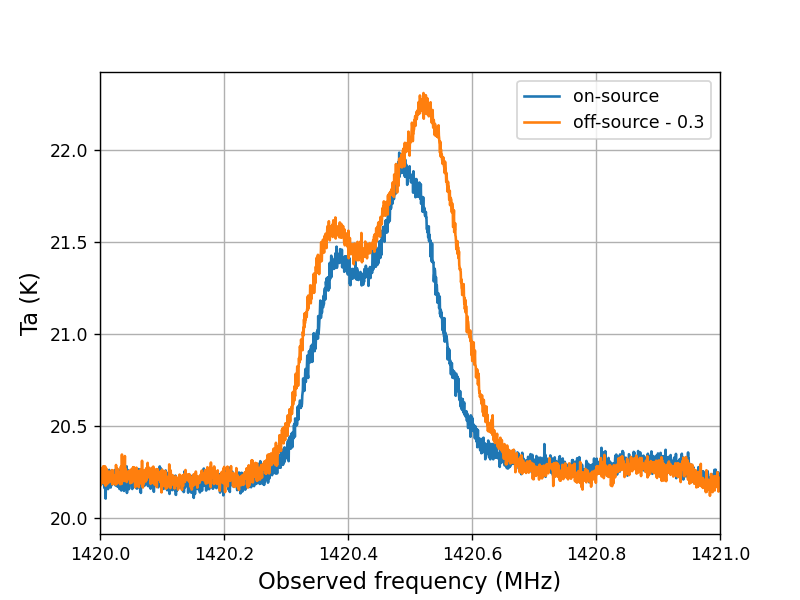}
\caption{The 20 minutes of integration of HI emission both for the on-source and off-source position during the observations of comet NEOWISE.
}
\label{fig.neowise_HI}
\end{figure}

For comet NEOWISE, although it was visually located in a clean background during its observations, there were still HI emissions around 1420.4 MHz both from the on-source position and off-position. The HI emission profile was less complicated than in the other scheduled observations of comets. Figure \ref{fig.neowise_HI} shows the 20 minutes of integration of the HI emission for the on-source position and off-position without Doppler correction from comet NEOWISE. Due to the different total on-source and off-source times for comet NEOWISE, we were unable to give a baseline subtraction from the off-source. It was also noted from Figure \ref{fig.neowise_HI} that off-source HI emission was slightly stronger than the on-source HI emission by less than 1 K. The calculated HI column density is 1.5 $\times$ 10$^{20}$ cm$^{-2}$ and 1.9 $\times$ 10$^{20}$ cm$^{-2}$ for the on-source and off-source positions, respectively. This difference could be due to the spatial variation of background HI gas. On the other hand, due to the large FAST beam size (174\arcsec) and small comet nucleus size (0.009\arcsec) at the observed distance, the obstruction of background emission by the comet could be ignored. Therefore, from these observations we could not constrain whether comet NEOWISE has an HI emission or not.

To our best knowledge, there are almost no report of HI in the radio wavelength range for comets to-date. Recently, \citet{Pal2024} reported an HI absorption detection from comet NEOWISE using the Giant Metrewave Radio Telescope (GMRT), and derived an HI column density on the order of 10$^{21}$ cm$^{-2}$. If we simply assume that HI and OH are mainly produced by the photodissociation of H$_2$O, then the production rate and column density for HI and OH should be similar. \citet{Keller1974} reported such a similar production rate for HI and OH with no less than 2 times difference for the comet Bennett (1970 II). For comet NEOWISE, compared with the derived OH column density of 1.1 $\times$ 10$^{13}$ cm$^{-2}$ from \citet{Smith2021}, the derived HI column density from \citet{Pal2024} was highly overestimated. From our observations, although the dip around 1420.4 MHz may be interpreted as an absorption feature from the on-source spectrum, we could not rule out the possibility of the multi-components of emission feature when compared with the off-source profile. Moreover, if we assume that the production rate for HI is comparable with that of OH, the integrated line intensity for HI would be below our detection limit.

\section{Discussion}\label{sec.dis}
Molecules in the coma originate from the outgassing of the nucleus, sublimation of ice species in the coma, or formed directly in the coma by photochemical processes \citep{Cordiner2023}. The chemical composition of comets is essential to study the chemical inheritance from the protosolar nebula to the planetary system. Both in-situ measurements and remote observations are used to identify the molecular species in comets. Numerous molecules have been discovered in Jupiter-family comet 67P/C-G, mainly due to the Rosetta mission of the European Space Agency. On the other hand, molecules in many long-period Oort Cloud comets have been observed during their perihelion passages by remote observations in the optical, infrared, or radio wavelengths. Table \ref{tab.mole} summarizes the detected molecules reported in the literature for the four comets we observed as derived from observations in various wavelengths.

The detectability of molecules depends on the activity of the comet. Molecules in highly active comets are relatively easy to detect \citep{Biver2015, Protopapa2021, Faggi2023}. Volatile species can be released during the sublimation of H$_2$O and CO$_2$ ices \citep{Rubin2023}, which is a consequence of the outburst due to the thermodynamic evolution of the cometary nucleus surface and subsurface layers \citep{Wesolowski2022}. Astrochemical models also show that abundant organic molecules can be present in the coma due to the outgassing of the nucleus and gas-phase chemistry in the coma \citep{Cordiner2021, Ahmed2022}. For the four comets that we observed, C/2020 R4 (ATLAS) was observed by only optical facilities. \citet{Lin2021} found at least three outbursts in their observations and did not see any new jet features and fragments based on its coma morphology. For comet NEOWISE and 67P/C-G, although multiple molecules have been observed in the frequency range higher than 100 GHz by IRAM 30-m radio telescope and NOEMA interferometry array, OH is the only molecule detected in the low-frequency L-band. Comet C/2021 A1 (Leonard) was observed by FAST about one month before its perihelion, although many volatile species were detected in the near-infrared range two weeks before the perihelion \citep{Faggi2023}.

In the previous section, the LTE condition was assumed to calculate the molecular production rates. However, in the cometary coma environment, the LTE condition may not be valid due to the processes of various other excitation mechanisms \citep{Bockelee-Morvan2004}. For example, although OH could be excited by the 2.7 K background radiation, the dominated mechanism to excite OH is the resonance fluorescence by the absorption of solar radiation \citep{Schleicher1988}. Therefore, the non-detection of radio wavelength of L-band molecular lines in the cometary environment could rule out the possibility of anomalously strong lines, which would be caused by pumping of ultraviolet/infrared radiation or other non-LTE effects.

In addition to the production rates of molecules, the detectability of molecules in the radio wavelengths depends on their intrinsic transitional characteristics and environmental excitation conditions, such as the Einstein A-coefficient and upper/lower energy at a specific level of excitation, as well as the gas temperature and number density of molecules in the coma. For the detection of more than 300 molecules in the ISM\footnote{https://cdms.astro.uni-koeln.de/classic/molecules}, the observed frequencies are in the range of tens to hundreds of GHz, and most of them are in star-forming regions \citep{McGuire2022}, suggesting that the excitation conditions of molecules are related to their environment. Although many molecular transitions fall in the L-band, including the organic and prebiotic molecules, the high upper-level energy (E$_{up}$) and/or low value of Einstein A-coefficient (A$_{ij}$) of the molecular excitation conditions do not favor their detection in the L-band in the cometary environment. For example, methanol, one of the molecules commonly detected in comets with high-frequency transitions, has E$_{up}$ and A$_{ij}$ generally several of tens Kelvin and higher than 10$^{-6}$ s$^{-1}$, respectively. However, the highest A-coefficient for molecules in the L-band is less than 10$^{-9}$ s$^{-1}$, and the E$_{up}$ can be as high as hundreds of Kelvin, making the detection much harder than in the high-frequency bands.

Finally, in the L-band, RFI is usually more severe than at high frequencies and has to be considered in data analysis. The wide frequency range of RFI reduces the usable frequency bandwidth and introduces ambiguities in the identifications of molecular spectral lines. The ultra-wide bandwidth receiver \citep{Zhang2023}, which will cover the 0.5-3.3 GHz frequency range, once installed on FAST, could greatly benefit the detection of molecules in comets.

\begin{table}
\begin{center}
\caption[]{Summary of the molecules detected by ground-based observations in the optical, infrared, and radio wavelengths for comets C/2020 F3 (NEOWISE), C/2020 R4 (ATLAS), 67P/C-G, and C/2021 A1 (Leonard).}
\label{tab.mole}
\begin{tabular}{p{90pt}p{50pt}p{180pt}p{50pt}}
\hline\noalign{\smallskip}
Comet               & Wavelength & Detected molecules & References \\
\hline\noalign{\smallskip}
C/2020 F3 (NEOWISE) & optical    & CN, CH, C$_2$, C$_3$, NH$_2$, CO$^+$, H$_2$O$^+$, N$_2^+$, CO, OCS, HCN, C$_2$H$_2$, NH$_3$, H$_2$CO, CH$_4$, C$_2$H$_6$, CH$_3$OH & 1, 2, 3, 4  \\
                    & infrared   & CN, NH$_2$, OH, H$_2$O, HCN, NH$_3$, CO, C$_2$H$_2$, C$_2$H$_6$, CH$_4$, CH$_3$OH, H$_2$CO & 5 \\
                    & radio      & OH, HCN, HNC, CH$_3$OH CS, H$_2$CO, CH$_3$CN, H$_2$S, CO & 6, 7, 8 \\
\hline
C/2020 R4 (ATLAS)   & optical    & CN, NH, C$_2$, C$_3$  & 9 \\
\hline
67P/C-G             & radio      & OH, HCN, CH$_3$OH, H$_2$S, CS, CH$_3$CN, H$_2$CO, HNCO, H$_2$O & 10 \\
\hline
C/2021 A1 (Leonard) & optical    & NH$_2$, C$_2$ & 11 \\
                    & infrared   & H$_2$O, HCN, NH$_3$, CO, C$_2$H$_2$, C$_2$H$_6$, CH$_4$, CH$_3$OH, H$_2$CO, OCS, HCl, CN, NH$_2$, OH  & 12  \\
                    & radio      & OH, HCN, HNC, CS, OCS, H$_2$S, H$_2$CO, HCOOH, HNCO, HC$_3$N, CH$_3$CN, NH$_2$CHO, CH$_2$CO, CH$_3$OH, CH$_3$CHO, CH$_2$OHCHO, C$_2$H$_5$OH, (CH$_2$OH)2   & 13, 14  \\
\hline\noalign{\smallskip}
\end{tabular}
\end{center}
\tablecomments{0.86\textwidth}{
1.~\citet{Cambianica2021a}, 2.~\citet{Cambianica2021b}, 3.~\citet{Faggi2020}, 4.~\citet{Munaretto2023},
5.~\citet{Faggi2021}, 6.~\citet{Smith2021}, 7.~\citet{Biver2022}, 8.~\citet{Drozdovskaya2023}, 9.~\citet{Manzini2021}, 10.~\citet{Biver2023}, 11.~\citet{Mugrauer2021}, 12.~\citet{Faggi2023}, 13.~\citet{Skirmante2022}, 14.~\citet{Biver2024}.}
\end{table}

\section{Summary}\label{sec.sum}
We observed four comets, C/2020 F3 (NEOWISE), C/2020 R4 (ATLAS), C/2021 A1 (Leonard), and 67P/Churyumov-Gerasimenko, to detect their molecular emission or absorption features in the radio L-band from 1.0 to 1.5 GHz using FAST from August 2020 to December 2021. We searched for thousands of line transitions associated with hundreds of molecular species in the RFI-free frequency channels in the data. No clear evidence of the emission lines was present, resulting in a null detection of those molecules in these four comets. Under the LTE conditions, we estimated the integrated intensity for CH$_3$OH at 1.443 GHz and $^{17}$OH at 1.302 GHz using their production rates reported in the literature, and found that the expected intensity for the searched molecular lines is too weak to be detected in our observations. Therefore, it is not surprising for the non-detection of molecular lines in the L-band from 1.0 to 1.5 GHz. This non-detection of molecular lines in the cometary environment could also rule out the possibility of unusually strong lines. Observing highly active comets and the implementation of the ultra-wide bandwidth receiver on FAST expected in the near future will improve the detectability of molecular lines in comets.

\begin{acknowledgements}
We thank the reviewer for the suggestions that helped us greatly improve the manuscript. The authors thank Qingliang Yang, Chun Sun at FAST Operation Center for assisting in the development of the special observation mode for moving target observations. The authors thank Zhong-Yi Lin and Wing-Huen Ip for their discussions on this project. This work is supported by a grant from the National Natural Science Foundation of China (NSFC) No.~11988101. L.-F.C. acknowledges the support from the China Postdoctoral Science Foundation grant No.~2023M733271 and the Foundation of Education Bureau of Guizhou Province, China (Grant No.~KY (2020) 003). Z.P. and L.Q. acknowledge the support from the National Key R\&D Program of China grant No.~2022YFC2205202 and 2020SKA0120100, and by the NSFC grant No.~11703047, 11773041, U2031119, 12173052, 12173053, 12373032, and 11963002. Z.P. and L.Q. are supported by the CAS ``Light of West China'' Program and the Youth Innovation Promotion Association of the Chinese Academy of Sciences (ID No.~2023064, 2018075, and Y2022027). D.L. is a New Cornerstone Investigator. D.Q. and X.-J.J. acknowledge the support by the NSFC grant No.~12373026.
\end{acknowledgements}

% \appendix                  %%appendicial material is supported

\bibliographystyle{raa}
\bibliography{ms2024-0130}

% \begin{thebibliography}{99}
%% you can type \apj for ApJ, \aap for A&A, \apss for Ap&SS, etc. Please consult
%% the macro chjaa.cls. You can also find them in aasguide.tex (AASTeX for ApJ, AJ, PASP)
%% Please follow the format of ChJAA's reference list

% \end{thebibliography}

\label{lastpage}

\end{document}